\documentclass[journal]{IEEEtran}

\usepackage{amsmath,amsfonts, amssymb}
\usepackage{cite}
\usepackage{algorithm, algpseudocode}
\usepackage{graphicx}
\usepackage{textcomp}
\usepackage[usenames,dvipsnames]{xcolor}
\usepackage{xspace}
\usepackage{fontawesome}
\usepackage{multirow}
\usepackage{rotating}
\usepackage{tikz}
\usepackage{soul}
\usepackage{makecell}
\usepackage{url}
\usepackage{enumitem}
\usepackage{booktabs}
\usepackage{fontawesome}
\usepackage{pifont}
\usepackage{tcolorbox}
\usepackage{tabularx}
\usepackage{array}
\usepackage{colortbl}
\usepackage{relsize}
\usepackage{listings}
\usepackage{etaremune}

\tcbuselibrary{skins}
\newcolumntype{Y}{>{\centering\arraybackslash}X}
\newcolumntype{L}{>{\raggedleft\arraybackslash}X}

\definecolor{darkGray}{gray}{0.15}
\definecolor{veryLightGray}{gray}{0.93}
\definecolor{lightGray}{gray}{0.85}
\tcbset{
  tab2/.style=
  {
    enhanced,
    fonttitle=\bfseries,fontupper=\normalsize\sffamily,
    colback=white, 
    colframe=black,
    colbacktitle=darkGray,
    coltitle=white,
    center title
  }
}

\tcbset{
  tab3/.style=
  {
    enhanced,
    fonttitle=\bfseries,fontupper=\normalsize\sffamily,
    colback=white, 
    colframe=black,
    colbacktitle=darkGray,
    coltitle=white,
    center title,
    boxsep=0pt,
    left=0pt,
    right=0pt,
    top=0pt,
    bottom=0pt
  }
}

\lstdefinelanguage{MoonBit}%
  {keywords={
      fn, let, mut, test, assert_eq, struct, enum, pub, trait, impl,
      typealias, for, while, if, else, in, is, match, match, guard,
      println, return
   },
   sensitive=true,
   morecomment=[l]{//},
   morestring=[s]{"}{"},
   morestring=[s]{'}{'},
   morekeywords=[2]{->,::,..=,..,>=,<=,==,!=,&&,||,+,-,*,/,\%,=},
}[keywords,comments,strings]

\newcommand{\lstbg}[3][0pt]{{\fboxsep#1\colorbox{#2}{\strut #3}}}
\lstdefinelanguage{diff}{
  basicstyle=\ttfamily\scriptsize,
  morecomment=[f][\lstbg{red!20}]-,
  morecomment=[f][\lstbg{green!20}]+,
  morecomment=[f][\textit]{@@},
}
    
\newcommand{\cmark}{\ding{51}}
\newcommand{\xmark}{\ding{55}}
\newcommand{\ie}{\emph{i.e.,}\xspace}
\newcommand{\eg}{\emph{e.g.,}\xspace}

\newcommand{\etal}{\emph{et~al.}\xspace}
\newcommand{\secref}[1]{Section~\ref{#1}\xspace}

\newcommand{\figref}[1]{Fig.~\ref{#1}\xspace}
\newcommand{\listref}[1]{Listing~\ref{#1}\xspace}
\newcommand{\tabref}[1]{Table~\ref{#1}\xspace}

\newcommand{\tse}[1]{{\color{black}#1}\xspace}

\newboolean{showcomments}
\setboolean{showcomments}{true}

\ifthenelse{\boolean{showcomments}}
  {\newcommand{\nb}[2]{
    \fbox{\bfseries\sffamily\scriptsize#1}
    {\sf\small$\blacktriangleright$\textit{#2}$\blacktriangleleft$}
   }
  }
  {\newcommand{\nb}[2]{}
  }

\begin{document}

\title{No Resource, No Benchmarks, No Problem? Evaluating and Improving LLMs for Code Generation in No-Resource Languages}

\author{Alessandro~Giagnorio,
	Alberto~Martin-Lopez,
	and~Gabriele~Bavota
\thanks{Alessandro Giagnorio and Gabriele Bavota are with SEART @ Software Institute, Universit\`a della Svizzera italiana. Alberto Martin-Lopez is with the SCORE Lab, I3US Institute, Universidad de Sevilla.}}

\markboth{}%
{Giagnorio \MakeLowercase{\textit{et al.}}: No-Resource, No Benchmarks, No Problem? Evaluating and Improving LLMs for Code Generation in No-Resource Languages}

\maketitle

\begin{abstract}
Large Language Models (LLMs) have significantly advanced the automation of software engineering tasks. One prominent example is code generation, where an LLM produces code in a specified programming language based on a natural language description.  Most research in this area has focused on high-resource languages, such as Python or Java, which benefit from abundant training data in repository platforms like GitHub. A smaller body of work has explored low-resource languages (\eg Lua, Racket), which are underrepresented in training corpora. In contrast, no-resource languages for which LLMs have seen virtually no training data remain largely unstudied. These languages often emerge in industry, where organizations develop proprietary or domain-specific languages unsupported by commercial tools like GitHub Copilot. This results in the need for companies to deploy their own in-house code recommenders. To investigate possible solutions in this context, we build and release three code generation benchmarks for no-resource languages, based on two recently proposed programming languages for which very little training data is available. Using these benchmarks, we experiment several solutions to teach LLMs about no-resource languages, including prompt-based techniques (\eg few-shot) as well as pre-training and fine-tuning exploiting the little data available. 
While further pre-training gives the largest performance gains for no-resource languages, applying it directly to instruction-tuned models harms their ability to follow instructions. To address this, we start from a base model, further pre-training it on the target language, and then inject instruction-following capabilities via weight diff transfer from an instruction model. Such an approach significantly improves code generation capabilities in no-resource settings, allowing companies to cheaply deploy an instruct model specialized on the  language of interest without dealing with the computational cost of instruction fine-tuning.
\end{abstract}

\begin{IEEEkeywords}
Code Generation, Benchmarks, Empirical Study, Large Language Models for Code
\end{IEEEkeywords}

\section{Introduction} \label{sec:introduction}

Code generation is the task of automatically implementing source code from a higher-level specification, typically written in natural language. The majority of existing work on Large Language Model (LLM)-based code generation has focused on ``high-resource programming languages,'' such as Python and Java \cite{reflexion,javabench,mapcoder}. These languages are well represented in public code repositories, making them dominant in the pre-training corpora of LLMs. As a result, models tend to perform particularly well on these languages \cite{cassano2023knowledge}. 

More recent studies have begun to investigate low-resource programming languages \cite{cassano:tse2023,cassano2023knowledge,giagnorio:icpc2025}, such as Lua, R, and Racket, which are characterized by relatively limited training data. While performance on these languages is generally lower than on high-resource ones, large-scale LLMs can still generalize reasonably well \cite{giagnorio:icpc2025}.

In contrast, little attention has been paid to no-resource programming languages, \ie languages that fall outside the pre-training distribution of LLMs. In particular, we focus on no-resource general-purpose languages, which share syntactic and semantic characteristics with mainstream programming languages but lack training data. As a consequence, commercial tools such as Copilot \cite{copilot} or ChatGPT \cite{chatgpt} do not support these languages, leaving organizations interested in AI-assisted programming with the challenge of developing custom, in-house solutions. Crafting effective and economically sustainable solutions is, however, far from trivial.

We make a number of contributions aimed at pushing forward research on code generation for \emph{no-resource programming languages}. We start by building and releasing three code generation benchmarks for this context. Benchmarks are used to assess the code generation performance of LLMs, and consist of a collection of coding tasks, each providing a natural language description (or specification) and a test suite aimed at assessing whether the LLM correctly implements the code. To the best of our knowledge, there are no publicly available benchmarks for no-resource languages since, as said, those are usually proprietary languages. To overcome this problem, we take as representative of no-resource languages Gleam \cite{gleam} and MoonBit \cite{moonbit}, two languages recently released and unlikely to be relevant in the training data of LLMs. Indeed, only very few repositories written in these languages can be found on GitHub (280 Gleam and 35 MoonBit repositories with $\geq$10 stars), with their popularity being extremely lower than those of languages considered in previous work as low-resource \cite{giagnorio:icpc2025} (\eg 18k R and 19k Lua GitHub repositories with $\geq$10 stars). To build benchmarks for these languages, we translated three code generation benchmarks, namely HumanEval \cite{humaneval}, MBPP \cite{mbpp}, and the subset of ``hard'' coding problems featured in McEval \cite{mceval}.
Having the benchmarks for \emph{no-resource languages}, we want to (i)~confirm the expected lack of support provided by modern LLMs, and (ii) explore techniques which would allow companies to come up with a working solution at reasonable cost. We run four state-of-the-art LLMs (\ie GPT-4o \cite{gpt4o}, o3-mini \cite{o3mini}, Qwen 2.5 Coder 32B Instruct \cite{qwen2.5coder32binstruct}, and Qwen 3 32B Instruct \cite{qwen3_32b}) on our benchmarks, obtaining---as expected---a very low $pass@1$ for both languages (\ie percentage of  tasks for which the LLM was able to produce a test-passing solution with 1 attempt). We also run the LLMs on the same three benchmarks but written in high-resource languages (Python and Java) and in languages considered by previous work as low-resource (Julia, Lua, R, Racket, Haskell). This was done to ``set the bar'' for what would be acceptable performance for a no-resource language. For example, if an LLM obtains $\sim$50\% $pass@1$ for low-resource languages, it is reasonable to see the 50\% as a sort of upper bound for its specialization to a no-resource language. To give an idea of the achieved results, we summarize the findings on the McEval-Hard benchmark, being the most challenging in our experiment: The four models achieve $pass@1$ scores being in the range of $\sim$59-89\% for high-resource languages (depending on the LLM and the language), 27-84\% for low-resource, and  0-1\% for no-resource.

Then, we experiment on the same four LLMs techniques aimed at boosting their performance on the no-resource languages. These include two in-context learning approaches, \ie few-shot and Retrieval-Augmented Generation (RAG), as well as pre-training and fine-tuning on the little data that can be collected for the two languages. The further pre-training (on top of the one which was already performed by the LLM's authors) was the best-performing technique, with the models approaching a $\sim$15\% $pass@1$ on the no-resource languages (McEval-Hard). However, the further pre-training can only be applied on ``base'' LLMs, namely their non-instruct models. Indeed, further pre-training an instruct model can degrade their instruction-following capabilities \cite{jindal2024balancing}, which is a key feature for an AI-based assistant. For this reason, we also experiment with an approach inspired by recent works in the Natural Language Processing (NLP)~\cite{jindal2024balancing, lin2025efficient} field that showed the possibility to transfer weights across LLMs having the same architecture. We thus start from a pre-trained base (non-instruct) model, M$_b$, for which an instruct version, M$_i$, is available. Then, we perform a further pre-training of M$_b$ exploiting the data available for the no-resource language, since this was the most effective technique we experimented with. At this point, M$_b$ acquires knowledge of the language~$k$ of interest  (M$_b$ $\rightarrow$ M$_{bk}$). However, since it is not an ``instruct'' model, M$_{bk}$ is not able to follow instructions. For this reason, we inject in it instruction-following capabilities by computing the weight diff between M$_i$ and M$_b$ and ``add'' such a diff to M$_{bk}$'s weights. This results in a model that features all instruction-following capabilities of an instruct model without incurring in the computational cost of full instruction fine-tuning. This approach further boosts the capabilities of the experimented models on no-resource languages, with $pass@1$ higher than 25\% on the McEval-Hard benchmark.

In summary, our contributions are: (i) three benchmarks (HumanEval, MBPP, and McEval-Hard) translated in two no-resource languages (Gleam and MoonBit); (ii) an empirical study showing the gap in code generation performance  LLMs experience when tested on high-, low-, and no-resource languages; and (iii) the experimentation of several techniques representing relatively cheap solutions allowing companies to deploy in-house coding assistants specialized for a language of interest. 
\vspace{-0.1cm}
\section{Study Design} \label{sec:design}

The \emph{goal} of the study is to experiment with techniques aimed at supporting the specialization of LLMs to no-resource languages. The \emph{context} consists of nine languages, including high-, low-, and no-resource, and six LLMs, including commercial and open models. We aim at answering the following research questions (RQs):

\textbf{RQ$_1$:} \emph{To what extent does the popularity of programming languages affect the code generation performance of LLMs?} This is a preliminary RQ, showing how the code generation performance of LLMs varies across languages characterized by a different amount of training data available on repositories such as GitHub. This will provide indications on (i) what are reasonable upper bounds expected for no-resource languages; and (ii) to what extent modern LLMs are able to cope with what have been considered in previous work as low-resource languages \cite{cassano2023knowledge,giagnorio:icpc2025}. In RQ$_1$, all LLMs are experimented in a zero-shot setting (\ie used out of the box).\vspace{0.1cm}

\textbf{RQ$_2$:} \emph{To what extent can in-context learning, pre-training, and fine-tuning boost the code generation performance of LLMs on no-resource languages?} In RQ$_2$ we assess whether few-shot, RAG, further pre-training and fine-tuning can help boosting the LLMs' performance on no-resource languages. Few-shot consists in prompting the LLM with concrete examples of code generations in the no-resource language of interest before asking for a new code generation task. RAG, instead, injects in the prompt information taken from the languages' documentation which is retrieved based on its relevance for the code generation task at hand. Concerning pre-training and fine-tuning, we use the little data available in GitHub repositories to teach the LLMs something about the no-resource languages.

\subsection{Context Selection}

In the following subsections we detail the context of our study in terms of (i) selected languages; (ii) code generation benchmarks; (iii) LLMs; and (iv) datasets used for pre-training and fine-tuning.

\subsubsection{Languages}

\textbf{No-Resource.} The main focus of our study is on two no-resource languages: Gleam and MoonBit. Gleam is a functional type-safe programming language designed for multi-thread scalability~\cite{gleam}. MoonBit is a general-purpose language designed for cloud and edge computing \cite{moonbit}. Both languages have been selected as representative of no-resource languages since their first stable versions have been proposed relatively recently, with Gleam v1 announced on 4 March 2024, and the MoonBit compiler made available on 18 December 2024. Note also that, being quite new, these languages are quickly evolving and, thus, even if recent LLMs may have seen some data about them during training, they may not have been exposed to the most recent languages' features (\eg MoonBit introduced a ``virtual packages'' feature on 16 May 2025). Also, their popularity on GitHub is extremely low compared to other languages, making them almost irrelevant in the LLMs' training sets. 

Despite the existence of several no-resource languages, we focus only on Gleam and MoonBit for three reasons: (i)~their stable release was launched after the cutoff date of the evaluated LLMs; (ii)~they are documented well enough for the authors to gain proper expertise for the creation of the benchmarks; and (iii)~they provide community support, to ask questions in case of doubts.
These criteria are important to ensure that the benchmarks we create for no-resource languages are of high quality and that the evaluated LLMs have likely not seen significant training data about them.

To give an idea of the different amounts of data available for the nine languages considered in our study, \tabref{tab:languages} shows their number of public GitHub repositories as collected on 2 July 2025. The classification as low- or high-resource languages follows previous studies in the literature \cite{cassano:tse2023,cassano2023knowledge,van:forge2024,giagnorio:icpc2025}, while the reported number of repositories per language helps in clarifying why Gleam and MoonBit can be considered as ``no-resource''. Indeed, they have at least one order of magnitude less GitHub repositories as compared to the least popular low-resource language (Racket). It is also important to consider that before March 2024 only 560 Gleam and 7 MoonBit GitHub repositories existed. This is a relevant date since, as we will explain later, it is the latest cutoff date for four of the LLMs considered in our study (\ie the LLM having the most recent training data has seen data up to March 2024). No official cutoff date is available for the remaining LLMs.

\begin{table}[t!]
	\centering
	\small
	\caption{Context selection: Programming languages.}
	\label{tab:languages}
	\begin{tabular}{llr}
		\toprule
		\textbf{Language} & \textbf{Classification} & \textbf{\#GitHub Repos.}\\
		\midrule
		MoonBit & no-resource & 400\\
		Gleam & no-resource & 2,900\\\midrule
		Racket & low-resource & 22,200\\
		Julia  & low-resource & 81,000\\
		Haskell  & low-resource & 155,000\\
		Lua & low-resource & 517,000\\
		R  & low-resource & 981,000\\\midrule
		Java & high-resource & 18,700,000\\
		Python & high-resource & 21,500,000\\ 
		\bottomrule
	\end{tabular}
\end{table}

\textbf{Low-Resource.} As done in recent studies  \cite{cassano:tse2023,cassano2023knowledge,van:forge2024,giagnorio:icpc2025}, we consider Racket, Julia, Haskell, Lua, and R as low-resource languages. As visible from \tabref{tab:languages}, the amount of potential training data they offer is substantially lower than that of high-resource languages. Still, within the set of low-resource languages we consider, there is a strong variability in their popularity, going from the $\sim$22k repositories of Racket up to the $\sim$981k of R. We study the extent to which their popularity impacts the LLMs' performance in RQ$_1$.

\textbf{High-Resource.} Java and Python are representative of high-resource languages in our study, with  $>$10M repositories each.

\subsubsection{Benchmarks}
To evaluate the code generation capabilities of the LLMs on the nine languages we resort to three benchmarks, namely HumanEval, MBPP, and McEval-Hard, with the last being a novel benchmark we propose. All these benchmarks exercise LLMs in function-level code generation (\ie given the description and signature of a function, finalize the implementation). We acknowledge that more complex and realistic benchmarks exist, like those tasking the LLMs with implementing the changes described in real issues (see \eg SWE-Bench \cite{jimenez2024swebench}). However, we decided to keep our focus on function-level benchmarks for mainly two reasons. First, our target on no-resource languages questions the ability of LLMs to cope with even self-contained and focused implementation tasks. Second, adopting more complex benchmarks would hinder a fair comparison across high-, low-, and no-resource languages, as it would require collecting semantically equivalent issues for multiple programming languages, a requirement that is difficult to satisfy in practice. In the following, we describe the three benchmarks.

\textbf{HumanEval and MBPP.} The first two have been taken from the MultiPL-E benchmark proposed by Cassano \etal \cite{cassano:tse2023}. MultiPL-E features the coding tasks of these two benchmarks translated to 24 programming languages, including all the high- and low-resource languages considered in our study. Each coding task represents a specific function to implement, described in natural language, and having associated tests to evaluate the correctness of the LLM's implementation.

To assess the performance of the LLMs on the no-resource languages, we also translated these two benchmarks in Gleam and MoonBit. The translation was performed by the first two authors starting from the Python version of the benchmarks and following a translation pipeline we defined. First, as done by Cassano \etal~\cite{cassano:tse2023}, we had to exclude 6 out of 161 tasks from HumanEval and 27 out of 399 in MBPP since they were too Python-specific and could not be translated into other languages. Also, a few problems were not available in MultiPL-E in all high- and low-resource languages subject of our study. Thus, to have a fair comparison among all languages, we considered only those available for all languages, which led to a final number of 154 problems for HumanEval and 355 for MBPP. For these coding tasks, we started by translating their prompts. Indeed, not only the function's signature is obviously different for each language, but also the textual description (docstring) requires translation. The problem arises from the fact that different languages use different terms to refer to the same coding construct. For example, an \texttt{Array} in Python corresponds to a \texttt{FixedArray} in MoonBit, while a Python's \texttt{List} corresponds to an \texttt{Array} in MoonBit. To have a starting point for the prompts translation, we instructed ChatGPT to implement these changes. 

This was done by explicitly providing ChatGPT with (i)~all textual transformations we expected (\eg \texttt{Array} $\rightarrow$ \texttt{FixedArray}) and (ii)~examples of good translations. The generated prompt translations were automatically checked for syntax errors in the signature and, then, manually inspected to find errors both in the signature and description. 
After obtaining the translated prompts, we proceeded to the translation of test suites. Once again, we started from translations proposed by ChatGPT: we provided the LLM with (i)~the original coding task, featuring both the original prompt and test suite; (ii)~the translated prompt manually checked; and (iii)~examples of good translations. The generated tests again represent a starting point and went through an automated syntax check and a subsequent manual inspection. The tests required major fixes due to ChatGPT's lack of knowledge of the two languages. 

\textbf{McEval-Hard.} The third benchmark considered in our study is McEval-Hard, that we built starting from McEval \cite{mceval}. McEval features coding tasks in 40 programming languages, including all the high- and low-resource ones considered in our study (but none of the no-resource). The coding tasks within each language are organized in three categories, based on their difficulty: easy, middle, hard. However, the coding tasks are not the same for the 40 languages, making a comparison of the LLMs' performance across the languages difficult. Also, there are only a few tasks per language ($\sim$50), out of which $<$10 are hard problems. To  build a challenging code generation benchmark being equal for all languages, we collected the ``hard tasks'' from McEval in each of the 40 languages (385 in total), filtered those without any tests (35 tasks), removed the 87 duplicated ones (\ie those present in more than one language) and those being too specific of the source language. This left us with 227 tasks that we translated in the nine languages subject of our study. The adopted translation pipeline is the same previously described for HumanEval and MBPP. 

Knowing that translating benchmarks can be error-prone, we also looked for the possibility of having the benchmarks for the no-resource languages double-checked by the creators of the languages themselves. We managed to have such a double check at least for Gleam (\url{https://lpil.uk/}) for a random sample of 50 instances: The feedback provided did not spot any major issue in our translation, but mostly recommended stylistic improvements which did not alter the code behavior, but made the code more compliant to the Gleam syntax.

\begin{table}[t!]
	\centering
	\small
	\caption{Context selection: LLMs.}
	\label{tab:llms}
	\resizebox{\columnwidth}{!}{%
	\begin{tabular}{l|rccc}
		\toprule
		& \textbf{\# Trainable} & \textbf{Instruction} & \textbf{Reasoning} & \textbf{Cutoff}\\
		& \textbf{\# Parameters} & \textbf{Following} & \textbf{Capabilities} & \textbf{Date}\\
		\midrule
		Qwen 2.5 Coder Base \cite{qwen2.5coder32b} & 32B & \xmark & \xmark & 2024-03 \\
		Qwen 2.5 Coder Instruct \cite{qwen2.5coder32binstruct} & 32B & \cmark & \xmark & 2024-03 \\
		Qwen 3 Base \cite{qwen3_8b_base} & 8B & \xmark & \xmark & ? \\ 
		Qwen 3 Instruct \cite{qwen3_32b} & 32B & \cmark & \cmark & ? \\
		o3-mini \cite{o3mini} & $\sim$200B & \cmark & \cmark & 2023-10\\
		GPT-4o \cite{gpt4o} & $\sim$200B & \cmark & \xmark & 2023-10\\
		\bottomrule
	\end{tabular}
	}
\end{table}

\subsubsection{LLMs}
\label{sub:llms}
We selected six LLMs diversifying between open and commercial, and with/without instruction-following and/or reasoning capabilities. \tabref{tab:llms} lists the selected LLMs, specifying their size, instruction-following and reasoning capabilities, and cutoff date, namely the date in which the collection of their training data has stopped. The latter information is not publicly available for all models (see question marks in the table). For the OpenAI models (\ie o3-mini and GPT-4o), the reported number of parameters is estimated, since such information is not publicly available.

In RQ$_1$ the four LLMs with instruction-following capabilities are experimented in zero-shot on all nine programming languages: Qwen 2.5 Coder 32B Instruct, Qwen 3 32B Instruct, o3-mini, and GPT-4o. In RQ$_2$, instead, we consider all of them, since different LLMs are suitable to experiment with different strategies to boost their performance on no-resource languages. In particular:\vspace{0.1cm}

\emph{In-context learning techniques (\ie few-shot and RAG)} are experimented on models having instruction-following capabilities: Qwen 2.5 Coder 32B Instruct, Qwen 3 32B Instruct, o3-mini, and GPT-4o.\vspace{0.05cm}

\eject

\emph{Pre-training} is experimented on non-instruct models, since further pre-training on instruct models is known to result in catastrophic forgetting of the instruction capabilities \cite{jindal2024balancing}:  Qwen 2.5 Coder 32B Base and Qwen 3 8B Base. Note that for Qwen 3 a 32B Base (\ie non Instruct) model is not available, thus explaining our choice of the 8B parameters here.\vspace{0.05cm}

\emph{Fine-tuning} is experimented on open instruct models, since it cannot be performed on closed models: Qwen 2.5 Coder 32B Instruct and Qwen 3 32B Instruct. 
\vspace{0.05cm}

\subsubsection{Datasets Used for Further Pre-Training and Fine-Tuning on Gleam and MoonBit}\label{sub:datasets}
We collected Gleam and MoonBit code from public GitHub repositories. Since these languages are both very recent, we defined a cut-off date to extract only up-to-date code: For Gleam we mined only repositories created after 5 March 2024 (\ie the day after the first stable version has been released). For MoonBit, instead, since we do not have an official date for the first release, we simply collected from all its repositories all files created in 2025, to ensure we capture the latest grammar and language features. This resulted in code files coming from a total of 2,159 Gleam and 262 MoonBit repositories.\vspace{0.05cm}

\textbf{Pre-Training.}
To avoid data leakage between the collected Gleam and MoonBit files and the used benchmarks, we perform a two-step process. First, we automatically remove all files containing a function having the same name of one of the functions in the benchmarks. Second, as done in recent work~\cite{muennighoff2025s1, openr1}, we extract all 8-grams composing each coding task featured in our benchmarks and check whether it is found in any of the pre-training files. In case of a match, the first author checked whether this was an actual case of data leakage or not, excluding the file from the pre-training dataset in the first case. At the end of this process, we obtained 18,767 Gleam and 3,609 MoonBit files for pre-training. In addition, we crawled the official documentation of the two languages from their respective websites. This documentation includes an overview of the languages, course materials, language cheat sheets, and descriptions of the standard libraries. The documentation is also part of the pre-training dataset.\vspace{0.05cm}

\textbf{Fine-Tuning.}
Starting from the code files in the pre-training dataset, we extracted all functions using the \texttt{tree-sitter}\footnote{\url{https://github.com/tree-sitter/tree-sitter}} library.
\tse{The \texttt{tree-sitter} parsers are developed by the very same Gleam\footnote{\url{https://github.com/gleam-lang/tree-sitter-gleam}} and MoonBit\footnote{\url{https://github.com/moonbitlang/tree-sitter-moonbit}} developers, thus we expect them to be reliable.} Based on the functions extracted, we filtered out those not having a docstring, with non-ASCII characters, or with a description shorter than 10 characters. Finally, we removed all functions having an empty body, TODO implementations, or being duplicated (\eg the same function is present across different pre-training files). Also, we performed the same two-step data leakage checks mentioned for the pre-training datasets, removing all suspect functions. This resulted in 13,534 Gleam and 2,444 MoonBit functions for the fine-tuning datasets. \tabref{tab:num-tokens} reports the total number of tokens in the pre-training and fine-tuning datasets, including both code and documentation tokens (computation done using the Qwen2.5 Coder tokenizer).

\begin{table}[t!]
	\centering
	\normalsize
	\caption{Number of tokens in Gleam and MoonBit datasets.}
	\label{tab:num-tokens}
	\resizebox{\columnwidth}{!}{%
	\begin{tabular}{llrrr}
		\toprule
		\textbf{Dataset} & \textbf{Language} & \textbf{\# Code Tokens}  & \textbf{\# Doc Tokens} & \textbf{\# Total Tokens}\\
		\midrule
		Pre-training & Gleam    & 28.2M     & 0.1M & 28.3M  \\
		  			 & MoonBit  & 13.1M   & 0.6M & 13.7M \\\midrule
		Fine-tuning  & Gleam    & 3.6M       & --- & 3.6M \\
		  			 & MoonBit  & 0.5M       & --- & 0.5M \\ 
		\bottomrule
	\end{tabular}
	}
\end{table}

\subsection{Data Collection}
\label{sub:collection}

\subsubsection{RQ$_1$}
In RQ$_1$ the LLMs are run in zero-shot on the three benchmarks of each of the nine languages using the prompts available in our replication package \cite{replication}. All prompts feature for each coding task the natural language description and the signature of the function to implement. We set the temperature to 0.2 when generating predictions for all models but o3-mini, which does not allow any temperature setting (thus, the default is used in this case). The temperature is used to control the randomness of the model's predictions, with 0 being the lowest and 2 the highest. The 0.2 setting is aligned with previous work in the literature (see \eg~\cite{humaneval,cassano:tse2023,giagnorio:icpc2025}). 

To account for the stochastic nature of LLMs, each model was run 10 times on each benchmark and language. In total, we performed 66,240 code generations with each model in zero-shot, for a total of 264,960 generations (66,240 $\times$ 4 models). 
All generations have been run against the respective test suites, to assess their correctness.

\subsubsection{RQ$_2$} 
We experiment with four strategies aimed at boosting LLMs' performance on the no-resource languages: \textit{few-shot}, \textit{retrieval-augmented generation} (RAG), \textit{pre-training}, and \textit{fine-tuning}. At inference time, we apply the same parameters used in RQ$_1$.

\textbf{Few-Shot.}
We use the fine-tuning dataset described in \secref{sub:datasets} as the \textit{knowledge base} from which to retrieve code examples. These examples are first transformed into embeddings using OpenAI's \texttt{text-embedding-3-large} model \cite{embedding}, and then indexed using the FAISS library \cite{douze2024faiss}. During code generation, we transform the benchmark prompt into embeddings and retrieve the top-five most similar examples from the processed \textit{knowledge base}. These examples are prepended to the final code generation prompt.

\textbf{RAG.}
We use the official documentation of the two no-resource languages as source for retrieval-augmented generation. Similarly to \textit{few-shot}, we transform the documentation into embeddings using the \texttt{text-embedding-3-large} model and store them in a vector database. Each vector is a ``subset'' of the language documentation (a paragraph) represented as an embedding. 

To retrieve the most relevant documentation for a given coding task, we follow a multi-step process guided by an LLM $M_e$, which in our implementation was \texttt{gpt-4o-mini-2024-07-18} \cite{gpt-4o-mini}:

\begin{enumerate}
	\item \textbf{Planning}: we prompt $M_e$ to generate a language-agnostic step-by-step plan for the coding task.
	\item \textbf{Query Generation}: for each step of the plan, we ask $M_e$ to generate a query that can be used to retrieve relevant documentation snippets from the vector database. For example, for a step like ``sort the list in input'', the query could be \textit{``How to sort a list?''}.
	
	\item \textbf{Retrieval}: for each query, we retrieve the five most relevant portions of documentation from the vector database and summarize them using $M_e$.
\end{enumerate}
Finally, we concatenate the generated queries and their summaries and use them to augment the task context.

\textbf{Pre-Training.}
We further pre-train the base versions of Qwen 2.5 Coder 32B and Qwen 3 8B on the pre-training datasets previously described. These datasets are chunked into sequences of 2,048 tokens, which we set as the maximum sequence length during training. Models are pre-trained using the Causal Language Modeling (CLM) objective and the \textit{LoRA} \cite{hu2022lora} technique. We adopt the same \textit{LoRA} hyperparameters used in \cite{weyssow2023exploring}, which are $r = 16$, $\alpha = 32$, and $dropout = 0.05$. The pre-training is performed for five epochs, using a learning rate of $5 \times 10^{-5}$, AdamW optimizer \cite{loshchilov:iclr19}, and a linear scheduler having a decaying factor of 0.

\textbf{Fine-Tuning.}
Similarly, we fine-tune the chosen LLMs with a maximum sequence length of 4,096 tokens on the fine-tuning datasets mentioned above. We use \textit{LoRA} with the same hyperparameters as for pre-training, and we train our models for five epochs using the recommended parameters from vendors, \ie learning rate of $5 \times 10^{-5}$, AdamW optimizer, and a cosine scheduler. For inference, we select the last epoch of each model, as the training loss converged in this epoch (see replication package \cite{replication}).

\subsection{Data Analysis}
The reference metric for evaluation and comparison is $pass@1$, where $1$ indicates the number of ``attempts'' a model is allowed to make. If the model's code passes \emph{all} unit tests for a given task, then $pass@1 = 1$; otherwise, $pass@1 = 0$. In addition, to capture partial correctness, we also report the percentage of passed unit tests ($passed_\%$), which provides a more fine-grained view of model performance in cases where not all tests are satisfied. For example, given a coding task having four tests and a candidate implementation provided by an LLM, we may have $pass@1 = 0$ and $passed_\%$ = 0.75 in the case in which the implementation results in three out of four tests passing. For the purpose of statistical significance, we run each LLM 10 times on each coding task in all RQs and compute both metrics with $n = 10$ repetitions.

In RQ$_1$, we use such metrics to compare the performance of the four LLMs across the nine languages, to observe the relationship between language popularity and LLMs' performance. In RQ$_2$ the focus is on the no-resource languages, and in particular on how few-shot, RAG, pre-training, and fine-tuning can boost performance as compared to a zero-shot setting. Besides showing the performance achieved by LLMs with the different strategies (zero-shot, few-shot, RAG, pre-training, fine-tuning), we also statistically compare these strategies against zero-shot, to assess whether the provided boost is significant. To make a concrete example, when contrasting the performance of o3-mini in zero-shot \emph{vs} few-shot on HumanEval-Gleam in terms of $pass@1$, we consider two distributions composed of 154 coding tasks $\times$ 10 repetitions = 1,540 $pass@1$ values. We use the McNemar's test~\cite{mcnemar}, which is suitable to do pairwise comparisons of dichotomous results of two different treatments. 

We adjust $p$-values using the Benjamini-Hochberg procedure~\cite{yoav:jstor1995} to account for multiple comparisons (\eg the performance of o3-mini in zero-shot is compared against what the same model achieves using few-shot, and RAG). We complement the McNemar's test with the Odds Ratio (OR) effect size to quantify the magnitude of the differences between the experimented methodologies.

\subsection{Replication Package}
We provide in our replication package \cite{replication}:
\begin{itemize}
\item The \emph{benchmarks} used in the form of JSONL files.
\item The model \emph{generations} collected across all RQs.
\item The \emph{prompts} used in the experiments.
\item The \emph{scripts} to replicate our experiments, from the data collection down to the implementation of all experimented techniques.
\item Additional \emph{results} discussed in the paper but not fully reported for the sake of brevity. 
\end{itemize}

\section{Results Discussion} \label{sec:results}

We discuss our findings by research question. Since our results are consistent between $pass@1$ and $passed_\%$, we only discuss the $pass@1$ findings, providing all data about $passed_\%$ in our replication package \cite{replication}.

\begin{table*}[t]
    \caption{LLMs performance ($pass@1$) on high-, low-, and no-resource programming languages.\vspace{-0.45cm}}
    \label{tab:low-resource}
    \small
\begin{tcolorbox}[tab3,title=LLMs comparison on different programming languages, boxrule=0.5pt]
    \relsize{-1}
\begin{tabularx}{\textwidth}{l!{\vrule width 1pt}Y|Y!{\vrule width 1pt}Y|Y|Y|Y|Y!{\vrule width 1pt}L|L}
\rowcolor{lightGray} \multicolumn{10}{c}{\textbf{GPT-4o}} \\\Xhline{1\arrayrulewidth}
 \textbf{Dataset} & \textbf{Python} & \textbf{Java} & \textbf{R} & \textbf{Lua} & \textbf{Haskell} & \textbf{Julia} & \textbf{Racket} & \makecell[c]{\textbf{Gleam}} & \makecell[c]{\textbf{MoonBit}} \\\Xhline{2\arrayrulewidth}
HumanEval   & 91.23 & 83.96 & 64.35 & 81.69 & 64.03 & 72.40 & 60.00 &  7.60 & 12.60 \\
MBPP        & 84.90 & 76.11 & 64.85 & 70.85 & 66.62 & 74.00 & 63.66 & 20.31 & 16.54 \\
McEval-Hard & 77.67 & 59.30 & 55.02 & 65.42 & 53.79 & 52.42 & 32.25 &  0.97 & 0.88 \\\Xhline{2\arrayrulewidth}

\rowcolor{lightGray} \multicolumn{10}{c}{\textbf{o3-mini}} \\\Xhline{1\arrayrulewidth}
 \textbf{Dataset} & \textbf{Python} & \textbf{Java} & \textbf{R} & \textbf{Lua} & \textbf{Haskell} & \textbf{Julia} & \textbf{Racket} & \makecell[c]{\textbf{Gleam}} & \makecell[c]{\textbf{MoonBit}} \\\Xhline{2\arrayrulewidth}
HumanEval   & 96.75 & 92.92 & 74.48 & 84.03 & 87.21 & 83.70 & 79.68 &  4.55 &  7.34 \\
MBPP        & 84.34 & 73.41 & 66.87 & 67.44 & 75.55 & 77.92 & 66.56 &  7.75 & 15.01 \\
McEval-Hard & 88.94 & 86.39 & 80.13 & 69.12 & 83.96 & 73.17 & 52.07 &  0.18 &  1.10 \\\Xhline{2\arrayrulewidth}

\rowcolor{lightGray} \multicolumn{10}{c}{\textbf{Qwen 2.5 Coder 32B Instruct}} \\\Xhline{1\arrayrulewidth}
 \textbf{Dataset} & \textbf{Python} & \textbf{Java} & \textbf{R} & \textbf{Lua} & \textbf{Haskell} & \textbf{Julia} & \textbf{Racket} & \makecell[c]{\textbf{Gleam}} & \makecell[c]{\textbf{MoonBit}} \\\Xhline{2\arrayrulewidth}
HumanEval   & 89.61 & 83.31 & 56.49 & 76.04 & 49.48 & 69.74 & 68.12 &  6.49 & 11.04 \\
MBPP        & 83.32 & 66.82 & 62.56 & 70.14 & 51.01 & 74.68 & 57.27 &  9.83 & 17.80 \\
McEval-Hard & 70.57 & 72.60 & 47.89 & 62.07 & 31.63 & 48.41 & 33.00 &  0.40 &  0.88 \\\Xhline{2\arrayrulewidth}

\rowcolor{lightGray} \multicolumn{10}{c}{\textbf{Qwen 3 32B Instruct}} \\\Xhline{1\arrayrulewidth}
 \textbf{Dataset} & \textbf{Python} & \textbf{Java} & \textbf{R} & \textbf{Lua} & \textbf{Haskell} & \textbf{Julia} & \textbf{Racket} & \makecell[c]{\textbf{Gleam}} & \makecell[c]{\textbf{MoonBit}} \\\Xhline{2\arrayrulewidth}
HumanEval   & 89.03 & 77.66 & 55.26 & 80.52 & 46.82 & 70.84 & 61.88 &  4.55 &  7.27 \\
MBPP        & 83.66 & 71.27 & 62.08 & 67.30 & 52.90 & 75.01 & 53.92 &  6.23 & 18.06 \\
McEval-Hard & 70.04 & 64.71 & 40.26 & 54.54 & 27.00 & 44.80 & 28.19 &  0.44 &  0.88 \\
\end{tabularx}
\end{tcolorbox}
\vspace{-0.3cm}
\end{table*}

\subsubsection*{\textbf{RQ$_1$: Effects of Language Popularity on LLMs' Code Generation Performance}}

\tabref{tab:low-resource} illustrates the performance of the four selected LLMs when used in a zero-shot setting (\ie out of the box) on high- (Python, Java), low- (R, Lua, Haskell, Julia, Racket), and no-resource (Gleam, MoonBit) programming languages, across all three benchmarks we experimented with. We discuss the results along two dimensions, namely language families and benchmarks. Languages are sorted from left to right based on their popularity on GitHub, as shown in \tabref{tab:languages}.

All studied LLMs exhibit high performance when evaluated on high-resource programming languages. The $pass@1$ obtained for these languages ranges between 59\% (Java, McEval-Hard, GPT-4o) and 97\% (Python, HumanEval, o3-mini), with an average performance of 79\% across all models and benchmarks. 

Results on low-resource languages, while more conservative, show in several cases performance comparable to those of high-resource languages. The $pass@1$ scores range between 27\% (Haskell, McEval-Hard, Qwen 3) and 87\% (Haskell, HumanEval, o3-mini), with an average of 62\%. In 49 out of 60 cases (5 languages $\times$ 3 benchmarks $\times$ 4 models), $pass@1$ is above 50\%. Even open-source models achieve over 50\% $pass@1$ in 20 out of 30 cases, indicating that low-resource languages may not be as challenging anymore as previously reported in the literature~\cite{cassano:tse2023, cassano2023knowledge, giagnorio:icpc2025} thanks to progress in LLMs' capabilities. Moreover, the distribution of results in \tabref{tab:low-resource} demonstrates that the performance of low-resource languages is not solely determined by their popularity. For instance, LLMs consistently perform better on Lua than on R, although the former has a smaller number of GitHub repositories compared to the latter (see \tabref{tab:languages}). As already observed in previous studies \cite{chen:icpc2022, giagnorio:icpc2025}, it is possible that the similarity with a high-resource language can help the model perform well on a low-resource language. For example, in this specific case, Lua is more similar to Python than R.

Results on no-resource languages are, as expected, radically different. Performance across LLMs and benchmarks stays between 0\% and 20\%, with an average of 9\%. The $pass@1$ scores $\geq$10\% achieved for Gleam and MoonBit on HumanEval and MBPP may appear surprising, but they can be explained by the fact that these benchmarks feature trivial coding tasks such as ``\emph{provide the sum of two integer numbers}'' or ``\emph{write a function to find the volume of a cube given its side length}''. 

\begin{lstlisting}[caption={Trivial task from MBPP benchmark for MoonBit.}, label={lst:trivial-mbpp}]
/// Write a function to find the volume of a
/// cube given its side length.
fn volume_cube(l: Int) -> Int {
    return l * l * l;
}
\end{lstlisting}

The latter case is depicted in \listref{lst:trivial-mbpp} for the MoonBit language and shows that such a coding task is not only trivial, but can be solved with knowledge of similar high-resource languages. Indeed, it is worth remembering that the LLM's prompt includes the function signature (\ie \texttt{fn volume\_cube(l: Int) -> Int \{\}}) which, thus, the LLM does not need to generate. The body ``\texttt{return l * l * l;}'' is common to many high-resource languages, such as Java. We also notice that the function's signature in MoonBit is quite similar to Rust ($\sim$1M repositories on GitHub) that, as Java, supports the ``\texttt{return l * l * l;}'' function implementation. Thus, LLMs may leverage their knowledge of similar languages to successfully generate completions for trivial coding tasks in no-resource languages. As it can be seen from \tabref{tab:low-resource}, as soon as the complexity of the coding task increases (our McEval-Hard benchmark), all LLMs consistently fail in no-resource languages (best $pass@1$ is 1\% by o3-mini on MoonBit).

We also performed an analysis of the reasons behind the LLMs' wrong code generations. In particular, we classified each wrong code generation as due to \emph{syntactic} or \emph{semantic} errors. The former identify code generations that violate the formal grammar of the target language (\ie there exists no parse tree / AST for it under that grammar). The latter, instead, are code generations that, while syntactically correct, result in either a runtime error or in at least a failing test. For the sake of brevity, full results are reported in our replication package \cite{replication}, while here we discuss the main findings.

We found a clear trend that differentiates no-resource languages from the high- and low-resource ones. Indeed, in Gleam and MoonBit the vast majority  of failures are due to syntactic errors, indicating that LLMs struggle with the basic syntax of these languages. For instance, for the best-performing model on these languages (\ie GPT-4o), roughly two-thirds of the failures are syntactic. This proportion is even higher for other models (\eg up to 90\% for o3-mini on Gleam). Differently, in the high- and low-resource languages, syntactic errors represent a minority of the failures (typically below 10\%), suggesting that LLMs generally possess a solid knowledge of their syntax. A notable exception is Java, which consistently exhibits a higher fraction of syntactic failures ($\sim$30\%). A plausible explanation for such a finding is Java's syntactic verbosity, which requires the correct placement of multiple mandatory elements (\eg class declarations, method signatures, types, and modifiers). Nevertheless, the overall success rate (\ie $pass@1$) for Java is substantially higher than for no-resource languages. Consequently, although the relative proportion of syntactic failures appears high for Java, the absolute number of syntactic errors is considerably smaller than for Gleam and MoonBit.

\begin{tcolorbox}[title={\faLightbulbO \hspace{0.1cm} Answer to RQ$_1$}, colframe=black!90, colback=gray!10, boxrule=0.8pt, arc=4pt, left=1pt, right=1pt, top=1pt, bottom=1pt]
    LLMs show excellent code generation performance on high-resource programming languages, with $pass@1$ scores close to 100\% in some cases. Low-resource languages show reasonable support with $pass@1$ scores above 50\% in most cases, with popularity not being the only determining factor. No-resource languages remain a challenge, with $pass@1$ scores below 20\% in most cases, and close to 0\% for hard coding tasks (McEval-Hard). Such a low performance is mainly due to the inability to produce syntactically-correct code for these languages, which leads to failures even for the most trivial tasks.
\end{tcolorbox}

\subsubsection*{\textbf{RQ$_2$: Boosting LLMs' Performance on No-Resource Languages via In-Context Learning, Pre-Training, and Fine-Tuning}}
We investigate techniques aimed at improving LLMs' performance on Gleam and MoonBit. 

\begin{table*}[t]
    \caption{LLMs performance on Gleam and MoonBit when using zero-shot, few-shot, RAG, pre-training, and fine-tuning.\vspace{-0.45cm}}
    \label{tab:no_resource_approach}
\small
\begin{tcolorbox}[tab3,title=Model performance with different settings on no-resource programming languages,boxrule=0.5pt]
\relsize{-1}

\setlength{\tabcolsep}{1.5pt}
\begin{tabularx}{\textwidth}{l!{\vrule width 1pt}r|r|r|r|r!{\vrule width 1pt}r|r|r|r|r}
\rowcolor{lightGray} \multicolumn{11}{c}{\textbf{GPT-4o}} \\\Xhline{1\arrayrulewidth}
\rowcolor{veryLightGray} \multicolumn{1}{c}{\textbf{}} & \multicolumn{5}{c}{\textbf{Gleam}}  & \multicolumn{5}{c}{\textbf{MoonBit}} \\\Xhline{1\arrayrulewidth}
 \textbf{Dataset} & \makecell[c]{\textbf{0-shot}} & \makecell[c]{\textbf{5-shot\phantom{a}}} & \makecell[c]{\textbf{RAG\phantom{a}}} & \makecell[c]{\textbf{Fine-tuned}} & \makecell[c]{\textbf{Pre-trained}} & \makecell[c]{\textbf{0-shot}} & \makecell[c]{\textbf{5-shot\phantom{a}}} & \makecell[c]{\textbf{RAG\phantom{a}}} & \makecell[c]{\textbf{Fine-tuned}} & \makecell[c]{\textbf{Pre-trained}} \\\Xhline{2\arrayrulewidth}
HumanEval    & 7.60 & 15.45 {\scriptsize(\texttt{~8.56})} & 14.22 {\scriptsize(\texttt{21.40})} & --- & --- & 12.60 & 25.97 {\scriptsize(\texttt{~5.29})} & 22.86 {\scriptsize(\texttt{~3.59})} & --- & --- \\
MBPP    & 20.31 & 30.37 {\scriptsize(\texttt{~3.19})} & 24.03 {\scriptsize(\texttt{~1.54})} & --- & --- & 16.54 & 40.51 {\scriptsize(\texttt{10.35})} & 32.68 {\scriptsize(\texttt{~4.43})} & --- & --- \\
McEval-Hard    & 0.97 & 1.23 {\scriptsize(\texttt{\textcolor{gray}{~1.32}})} & 2.16 {\scriptsize(\texttt{~3.45})} & --- & --- & 0.88 & 8.06 {\scriptsize(\texttt{17.30})} & 6.12 {\scriptsize(\texttt{12.90})} & --- & --- \\
\Xhline{2\arrayrulewidth}
\rowcolor{lightGray} \multicolumn{11}{c}{\textbf{o3-mini}} \\\Xhline{1\arrayrulewidth}
\rowcolor{veryLightGray} \multicolumn{1}{c}{\textbf{}} & \multicolumn{5}{c}{\textbf{Gleam}}  & \multicolumn{5}{c}{\textbf{MoonBit}} \\\Xhline{1\arrayrulewidth}
 \textbf{Dataset} & \makecell[c]{\textbf{0-shot}} & \makecell[c]{\textbf{5-shot\phantom{a}}} & \makecell[c]{\textbf{RAG\phantom{a}}} & \makecell[c]{\textbf{Fine-tuned}} & \makecell[c]{\textbf{Pre-trained}} & \makecell[c]{\textbf{0-shot}} & \makecell[c]{\textbf{5-shot\phantom{a}}} & \makecell[c]{\textbf{RAG\phantom{a}}} & \makecell[c]{\textbf{Fine-tuned}} & \makecell[c]{\textbf{Pre-trained}} \\\Xhline{2\arrayrulewidth}
HumanEval    & 4.55 & 9.74 {\scriptsize(\texttt{17.00})} & 9.29 {\scriptsize(\texttt{~8.30})} & --- & --- & 7.34 & 39.22 {\scriptsize(\texttt{31.69})} & 32.08 {\scriptsize(\texttt{10.07})} & --- & --- \\
MBPP    & 7.75 & 19.66 {\scriptsize(\texttt{~6.35})} & 18.87 {\scriptsize(\texttt{~5.76})} & --- & --- & 15.01 & 45.92 {\scriptsize(\texttt{23.85})} & 40.79 {\scriptsize(\texttt{~8.32})} & --- & --- \\
McEval-Hard    & 0.18 & 0.93 {\scriptsize(\texttt{~6.67})} & 0.57 {\scriptsize(\texttt{~5.50})} & --- & --- & 1.10 & 12.20 {\scriptsize(\texttt{19.00})} & 9.65 {\scriptsize(\texttt{13.12})} & --- & --- \\
\Xhline{2\arrayrulewidth}
\rowcolor{lightGray} \multicolumn{11}{c}{\textbf{Qwen 2.5 Coder 32B Instruct}} \\\Xhline{1\arrayrulewidth}
\rowcolor{veryLightGray} \multicolumn{1}{c}{\textbf{}} & \multicolumn{5}{c}{\textbf{Gleam}}  & \multicolumn{5}{c}{\textbf{MoonBit}} \\\Xhline{1\arrayrulewidth}
 \textbf{Dataset} & \makecell[c]{\textbf{0-shot}} & \makecell[c]{\textbf{5-shot\phantom{a}}} & \makecell[c]{\textbf{RAG\phantom{a}}} & \makecell[c]{\textbf{Fine-tuned}} & \makecell[c]{\textbf{Pre-trained}} & \makecell[c]{\textbf{0-shot}} & \makecell[c]{\textbf{5-shot\phantom{a}}} & \makecell[c]{\textbf{RAG\phantom{a}}} & \makecell[c]{\textbf{Fine-tuned}} & \makecell[c]{\textbf{Pre-trained}} \\\Xhline{2\arrayrulewidth}
HumanEval    & 6.49 & 7.79 {\scriptsize(\texttt{~3.00})} & 14.29 {\scriptsize(\texttt{13.00})} & 24.74 {\scriptsize(\texttt{32.22})} & --- & 11.04 & 23.25 {\scriptsize(\texttt{~9.17})} & 27.08 {\scriptsize(\texttt{~4.48})} & 34.74 {\scriptsize(\texttt{11.43})} & --- \\
MBPP    & 9.83 & 18.87 {\scriptsize(\texttt{~4.21})} & 21.44 {\scriptsize(\texttt{~5.58})} & 34.03 {\scriptsize(\texttt{14.85})} & --- & 17.80 & 33.86 {\scriptsize(\texttt{~4.08})} & 30.48 {\scriptsize(\texttt{~3.03})} & 37.38 {\scriptsize(\texttt{~5.34})} & --- \\
McEval-Hard    & 0.40 & 1.32 {\scriptsize(\texttt{~3.33})} & 0.88 {\scriptsize(\texttt{23.00})} & 3.04 {\scriptsize(\texttt{121.0})} & --- & 0.88 & 5.81 {\scriptsize(\texttt{57.00})} & 7.05 {\scriptsize(\texttt{15.00})} & 10.93 {\scriptsize(\texttt{457.0})} & --- \\
\Xhline{2\arrayrulewidth}
\rowcolor{lightGray} \multicolumn{11}{c}{\textbf{Qwen 3 32B Instruct}} \\\Xhline{1\arrayrulewidth}
\rowcolor{veryLightGray} \multicolumn{1}{c}{\textbf{}} & \multicolumn{5}{c}{\textbf{Gleam}}  & \multicolumn{5}{c}{\textbf{MoonBit}} \\\Xhline{1\arrayrulewidth}
 \textbf{Dataset} & \makecell[c]{\textbf{0-shot}} & \makecell[c]{\textbf{5-shot\phantom{a}}} & \makecell[c]{\textbf{RAG\phantom{a}}} & \makecell[c]{\textbf{Fine-tuned}} & \makecell[c]{\textbf{Pre-trained}} & \makecell[c]{\textbf{0-shot}} & \makecell[c]{\textbf{5-shot\phantom{a}}} & \makecell[c]{\textbf{RAG\phantom{a}}} & \makecell[c]{\textbf{Fine-tuned}} & \makecell[c]{\textbf{Pre-trained}} \\\Xhline{2\arrayrulewidth}
HumanEval    & 4.55 & 9.09 {\scriptsize(\texttt{141.0})} & 10.39 {\scriptsize(\texttt{181.0})} & 23.57 {\scriptsize(\texttt{30.30})} & --- & 7.27 & 22.40 {\scriptsize(\texttt{~7.30})} & 23.96 {\scriptsize(\texttt{~6.98})} & 34.81 {\scriptsize(\texttt{13.85})} & --- \\
MBPP    & 6.23 & 23.94 {\scriptsize(\texttt{16.34})} & 19.46 {\scriptsize(\texttt{11.22})} & 37.32 {\scriptsize(\texttt{37.80})} & --- & 18.06 & 32.59 {\scriptsize(\texttt{~6.21})} & 32.48 {\scriptsize(\texttt{~3.81})} & 41.94 {\scriptsize(\texttt{12.94})} & --- \\
McEval-Hard    & 0.44 & 0.88 {\scriptsize(\texttt{\textcolor{gray}{~2.00}})} & 1.32 {\scriptsize(\texttt{~3.00})} & 3.88 {\scriptsize(\texttt{~8.80})} & --- & 0.88 & 5.46 {\scriptsize(\texttt{11.40})} & 6.48 {\scriptsize(\texttt{255.0})} & 13.04 {\scriptsize(\texttt{553.0})} & --- \\
\Xhline{2\arrayrulewidth}
\rowcolor{lightGray} \multicolumn{11}{c}{\textbf{Qwen 2.5 Coder 32B Base}} \\\Xhline{1\arrayrulewidth}
\rowcolor{veryLightGray} \multicolumn{1}{c}{\textbf{}} & \multicolumn{5}{c}{\textbf{Gleam}}  & \multicolumn{5}{c}{\textbf{MoonBit}} \\\Xhline{1\arrayrulewidth}
 \textbf{Dataset} & \makecell[c]{\textbf{0-shot}} & \makecell[c]{\textbf{5-shot\phantom{a}}} & \makecell[c]{\textbf{RAG\phantom{a}}} & \makecell[c]{\textbf{Fine-tuned}} & \makecell[c]{\textbf{Pre-trained}} & \makecell[c]{\textbf{0-shot}} & \makecell[c]{\textbf{5-shot\phantom{a}}} & \makecell[c]{\textbf{RAG\phantom{a}}} & \makecell[c]{\textbf{Fine-tuned}} & \makecell[c]{\textbf{Pre-trained}} \\\Xhline{2\arrayrulewidth}
HumanEval    & 1.95 & --- & --- & --- & 32.99 {\scriptsize(\texttt{957.0})} & 7.14 & --- & --- & --- & 41.62 {\scriptsize(\texttt{18.70})} \\
MBPP    & 3.46 & --- & --- & --- & 47.35 {\scriptsize(\texttt{120.8})} & 12.42 & --- & --- & --- & 44.76 {\scriptsize(\texttt{~9.63})} \\
McEval-Hard    & 0.00 & --- & --- & --- & 12.47*  & 1.67 & --- & --- & --- & 25.86 {\scriptsize(\texttt{35.31})} \\
\Xhline{2\arrayrulewidth}
\rowcolor{lightGray} \multicolumn{11}{c}{\textbf{Qwen 3 8B Base}} \\\Xhline{1\arrayrulewidth}
\rowcolor{veryLightGray} \multicolumn{1}{c}{\textbf{}} & \multicolumn{5}{c}{\textbf{Gleam}}  & \multicolumn{5}{c}{\textbf{MoonBit}} \\\Xhline{1\arrayrulewidth}
 \textbf{Dataset} & \makecell[c]{\textbf{0-shot}} & \makecell[c]{\textbf{5-shot\phantom{a}}} & \makecell[c]{\textbf{RAG\phantom{a}}} & \makecell[c]{\textbf{Fine-tuned}} & \makecell[c]{\textbf{Pre-trained}} & \makecell[c]{\textbf{0-shot}} & \makecell[c]{\textbf{5-shot\phantom{a}}} & \makecell[c]{\textbf{RAG\phantom{a}}} & \makecell[c]{\textbf{Fine-tuned}} & \makecell[c]{\textbf{Pre-trained}} \\\Xhline{2\arrayrulewidth}
HumanEval    & 1.62 & --- & --- & --- & 18.57 {\scriptsize(\texttt{19.64})} & 8.18 & --- & --- & --- & 36.82 {\scriptsize(\texttt{18.64})} \\
MBPP    & 4.65 & --- & --- & --- & 23.63 {\scriptsize(\texttt{22.06})} & 12.79 & --- & --- & --- & 42.08 {\scriptsize(\texttt{10.12})} \\
McEval-Hard    & 0.35 & --- & --- & --- & 4.36 {\scriptsize(\texttt{12.38})} & 0.44 & --- & --- & --- & 19.87 {\scriptsize(\texttt{50.00})} \\
\end{tabularx}
\end{tcolorbox}
\vspace{-0.1cm} \footnotesize *OR not computable since the LLM in 0-shot achieves 0.00 $pass@1$.
\vspace{-0.4cm}
\end{table*}

\tabref{tab:no_resource_approach} presents the $pass@1$ on the three benchmarks of the six LLMs subject of this RQ using zero-shot (as in RQ$_1$) plus four new strategies: \textit{5-shot}, \emph{RAG}, \textit{fine-tuning}, and \textit{pre-training}. As explained in \secref{sub:llms}, not all techniques have been experimented on all LLMs: \textit{few-shot} and \textit{RAG} are experimented on all models having instruction-following capabilities (GPT-4o, o3-mini, Qwen 2.5 Coder 32B Instruct, and Qwen 3 32B Instruct);  \emph{fine-tuning} on open instruct models (Qwen 2.5 Coder 32B Instruct and Qwen 3 32B Instruct); and \emph{pre-training} on non-instruct models only (Qwen 2.5 Coder 32B Base and Qwen 3 8B Base).

Each $pass@1$ value in \tabref{tab:no_resource_approach} is accompanied by the Odds Ratio (OR) output of the McNemar's test comparing the performance of an LLM$_i$ with a given technique (\eg few-shot) versus the performance of LLM$_i$ used in zero-shot (baseline). For example, applying few-shot to GPT-4o on Gleam, increases the $pass@1$ on HumanEval from 7.60\% to 15.45\%, resulting in an OR=8.56, which indicates that, among the discordant cases, few-shot was over 8 times more likely than zero-shot to produce a correct implementation.
The few ORs that are not statistically significant are reported in grey in \tabref{tab:no_resource_approach}. Note that a higher OR does not always imply a higher gap in $pass@1$. Indeed, McNemar's OR quantifies the imbalance in discordant outcomes (\ie coding tasks for which one technique produces a correct output and the other does not). For example, a higher OR may arise from a small number of discordant cases with a strong directional imbalance, while a lower OR might come from a larger number of discordant cases that are more balanced. 

Starting from in-context learning techniques (\ie 5-shot and RAG), we observe that few-shot is slightly more effective than RAG. Indeed, in 7 out of 12 cases for Gleam and 8 out of 12 for MoonBit, few-shot outperforms RAG in terms of $pass@1$. We hypothesize that models are better able at grasping the grammar of unfamiliar languages from code examples rather than from relevant portions of the documentation, which may be more or less code-oriented depending on the language. This is partially confirmed by the fact that few-shot reduces syntax errors by 15.36\% compared to zero-shot, while RAG achieves a smaller reduction of 8.94\%. 

It can also be seen from \tabref{tab:no_resource_approach} that the boost in performance provided by in-context learning techniques is benchmark- and language-dependent. Indeed, such an improvement is higher (i)~on MoonBit than on Gleam; and (ii) on simpler coding tasks (HumanEval and MBPP) than on more complex ones (McEval-Hard). The higher gain on MoonBit than on Gleam can be explained by two factors. First, as shown in \tabref{tab:languages}, MoonBit has seven times less GitHub repositories than Gleam. Thus, for this language, we can conjecture that the additional information about the language provided via in-context learning is likely to make a stronger difference. Second, as stated in the paper presenting MoonBit \cite{moonbit}, this language has been designed to be AI-friendly, with an AI-driven language design (see Section 2 in \cite{moonbit}) also featuring aspects of the language allowing a ``\emph{more flexible retrieval-based prompt augmentation}'' \cite{moonbit}.

As per the coding task complexity, it can be seen that on McEval-Hard the gain in performance is substantially lower for both languages, suggesting that showing examples (few-shot) or relevant parts of the documentation (RAG) in the prompt is not enough when dealing with challenging programming tasks.

Focusing on training-based approaches, the fine-tuned Qwen 2.5 Coder 32B Instruct and Qwen 3 32B Instruct outperform zero-shot and in-context learning techniques applied on the same models. 
Notably, fine-tuned open-source models often outperform commercial models in their best setting, especially on Gleam. For example, the Gleam fine-tuned version of Qwen 3 32B Instruct achieved 23.57\% on HumanEval, 37.32\% on MBPP, and 3.88\% on McEval-Hard, which can be compared against the best LLM with in-context learning (\ie GPT-4o with 5-shot) which achieved on the same three benchmarks 15.45\%, 30.37\%, and 1.23\%, respectively. On MoonBit, instead, the best LLM using in-context learning is o3-mini which performs slightly better than the fine-tuned Qwen 3 32B Instruct (see \tabref{tab:no_resource_approach}). In summary, given the same models, fine-tuning is superior to in-context learning. Also, fine-tuned open models are competitive with commercial ones used with few-shot or RAG.
 
The last technique we analyze is the further pre-training of the base models. Let us start from the results achieved on Qwen 2.5 Coder 32B Base, which can be compared against what we observed for its ``instruct'' version. For the latter, fine-tuning was by far the best-performing technique. Thus, we use it as comparison against its pre-trained base version. On all benchmarks and for both languages, the pre-trained base model is superior to the fine-tuned instruct model. Remember that we are comparing two models having the same size and the same architecture, with the instruct version just being a further instruction-tuned version of the base model. 
The gap in performance is major in all cases. For Gleam: on HumanEval, 32.99 (pre-trained) \emph{vs} 24.74 (fine-tuned); on MBPP, 47.35 \emph{vs} 34.03; and on McEval-Hard, 12.47 \emph{vs} 3.04. For MoonBit: on HumanEval, 41.62 \emph{vs} 34.74; on MBPP, 44.76 \emph{vs} 37.38; and on McEval-Hard, 25.86 \emph{vs} 10.93. All these differences are statistically significant (McNemar test, $p$-values $<$ 0.05 after Benjamini-Hochberg correction), with ORs ranging from 1.89 to 4.24 for Gleam and from 1.43 to 3.78 for MoonBit.

When looking at the pre-trained Qwen 3 8B Base, in this case we do not have an identical model to compare with. However, when looking at the results, we can see that on MoonBit the pre-trained 8B model has comparable performance to the fine-tuned 32B model: on HumanEval, 36.82 (pre-trained) \emph{vs} 34.81 (fine-tuned); on MBPP, 42.08 \emph{vs} 41.94; and on McEval-Hard, 19.87 \emph{vs} 13.04. On Gleam, instead, the larger fine-tuned model is superior on HumanEval and MBPP, while worst on McEval-Hard. These differences are statistically significant in 1 case in favor of the 8B pre-trained model, and in 2 cases of the 32B fine-tuned model, not showing a clear winner.

Putting all above-discussed evidence together, we conclude that a further pre-training helps more than fine-tuning for no-resource languages. This can be explained by the different amount of data that can be exploited in the two training scenarios, as visible from \tabref{tab:num-tokens}. Indeed, when pre-training, the entire code files as well as any source of language documentation can be used for teaching the language to the model. Instead, fine-tuning requires the building of natural language descriptions of the code to implement paired with a corresponding code implementation. In our setting (\ie function-level), this means mining from the very few repositories available for the no-resource language only the functions having a non-empty description, excluding everything else. This is the reason behind the much larger amount of training tokens available in the pre-training datasets (28.3M for Gleam, 13.7M for MoonBit) as compared to the fine-tuning datasets (3.6M for Gleam, 0.5M for MoonBit).

Similarly to what done in RQ$_1$, we looked at the impact of in-context learning, pre-training, and fine-tuning on the reduction of \emph{syntactic} and \emph{semantic} errors (full table in our replication package \cite{replication}). Within in-context learning, few-shot prompting is consistently more effective than RAG at reducing syntactic errors. This result holds across both languages and all four LLMs evaluated. For Gleam, even with few-shot learning syntactic errors remain the predominant cause of failure (over semantic ones) across all LLMs. In contrast, this pattern does not hold for MoonBit. Specifically, on the two Qwen models, $\sim$50\% of the failures are due to syntactic errors, while for the GPT-based models semantic errors represent roughly two-thirds of the overall failures. This again suggests a higher effectiveness of few-shot learning on MoonBit for the reasons previously explained, with LLMs getting a better understanding of the language syntax. 

As expected, fine-tuning and pre-training lead to a substantial reduction in syntactic errors. Both approaches shift LLM behavior toward what we observed for high- and low-resource languages, with fewer than 20\% of failures attributable to syntactic mistakes across all LLM-language combinations. These results confirm the superior effectiveness of training-based approaches compared to in-context learning methods for teaching no-resource languages to LLMs.

\begin{tcolorbox}[title={\faLightbulbO \hspace{0.1cm} Answer to RQ$_2$}, colframe=black!90, colback=gray!10, boxrule=0.8pt, arc=4pt, left=1pt, right=1pt, top=1pt, bottom=1pt]
    In-context learning techniques help in improving code generation performance for no-resource languages as compared to zero-shot. However, training-based techniques (fine-tuning and pre-training) are more effective than in-context learning, allowing open models to achieve performance superior to commercial LLMs. Pre-training is the most promising technique since, differently from fine-tuning, allows to exploit ``all'' little data available for the no-resource language, achieving the strongest boost in performance among all experimented techniques. On the negative side, the pre-trained base models do not have instruction-following capabilities, being thus suboptimal as AI coding assistants. In \secref{sec:approach} we address this limitation.
\end{tcolorbox}
\section{Instruction Transferring} \label{sec:approach}

In our answer to RQ$_2$, we showed that the best approach to boost performance on no-resource languages is further pre-training a base model on the available data, even if scarce. However, the resulting model does not have instruction-following capabilities, which are crucial for an AI coding assistant. Indeed, prompts (\ie natural language descriptions of the desired code) can vary widely in form. 
This variability is already evident in existing benchmarks, which often use diverse prompting styles, but it becomes even more pronounced in real-world scenarios, where different developers may express the same request in very different ways.

To address this limitation, one possible solution would be to perform an additional \emph{instruction fine-tuning} on top of the further pre-trained base model. However, (i) instruction-tuning datasets are typically not publicly available, and (ii) the cost of such a process is known to be extremely high~\cite{hu2022lora}, since  large datasets are needed. 

As an alternative approach, researchers in the Natural Language Processing (NLP) community recently suggested \emph{fine-tuning reuse}~\cite{jindal2024balancing, lin2025efficient}, which allows to transfer the instruction-following capabilities of a model $M_i$ to another \emph{base} model (\ie without instruction-following capabilities). In particular, such an application assumes the existence of three models all having the same size and architecture: (i) $M_i$, the one with instruction-following capabilities; (ii) $M_b$, the base model on top of which $M_i$ has been created via instruction fine-tuning; and (iii) $M_{bk}$, a version of $M_b$ further trained to better support a specific task or language of interest. By computing the diff between $M_i$'s and $M_b$'s weights ($\Delta_{w}$), we can capture ``the portion of $M_i$'s knowledge'' allowing it to follow complex instructions. We can then sum $\Delta_{w}$ to $M_{bk}$'s weights, obtaining---at a negligible cost---a new instruct model ($M_{bk+i}$), which is specialized on the task/language of interest.\footnote{Cost is negligible since the diff between models can be computed using CPUs only.}

We experiment this approach as a further attempt to boost the performance of code models on no-resource languages. In particular, we answer the following research question:\vspace{0.05cm}

\textbf{RQ$_3$:} \emph{To what extent does instruction transferring boost the code generation performance of LLMs on no-resource languages?} Our $M_{bk}$ are base models further pre-trained on the no-resource language of interest (as done in RQ$_2$), while $M_i$ and $M_b$ are the instruct and base versions of that same model, as  released by their authors. 

We use as $M_i$ models the already mentioned Qwen 2.5 Coder 32B Instruct and Qwen 3 8B Instruct. The latter is a reasoning model, thus the weighting diff in this case is expected to inject reasoning capabilities into the $M_{bk}$ models. For both models we have their $M_b$ versions available (\ie Qwen 2.5 Coder 32B Base and Qwen 3 8B Base), thus allowing the computation of the weights diff $\Delta_{w}$. Finally, our $M_{bk}$ models are the versions of Qwen 2.5 Coder 32B Base and Qwen 3 8B Base further pre-trained on Gleam and MoonBit.

\begin{table}[thb]
    \vspace{-0.25cm}
    \caption{LLMs performance when using base + pre-training (PT) and base + pre-training + instruction transferring (Diff).\vspace{-0.35cm}}
    \label{tab:diffResults}
    \footnotesize
    \begin{tcolorbox}[tab3, 
        title=Difference between pre-training and instruction transferring,
        boxrule=0.5pt]
    \relsize{-1}

    \setlength{\tabcolsep}{1.5pt}
    \begin{tabularx}{\columnwidth}{l!{\vrule width 1pt}L|L|L!{\vrule width 1pt}L|L|L}
        \rowcolor{lightGray}
        \multicolumn{7}{c}{\textbf{Qwen 2.5 Coder 32B}} \\
        \Xhline{1\arrayrulewidth}
        \rowcolor{veryLightGray}
        \textbf{} &
        \multicolumn{3}{c!{\vrule width 1pt}}{\textbf{Gleam}} &
        \multicolumn{3}{c}{\textbf{MoonBit}} \\
        \Xhline{1\arrayrulewidth}
        \textbf{Benchmark} & \makecell[c]{\textbf{PT}} & \makecell[c]{\textbf{Diff}} & \makecell[c]{$\Delta$} & \makecell[c]{\textbf{PT}} & \makecell[c]{\textbf{Diff}} & \makecell[c]{$\Delta$} \\
        \Xhline{2\arrayrulewidth}
        HumanEval    & 32.99 & 56.23 & \raisebox{0.1ex}{\color{ForestGreen}{$\blacktriangle$}}~23.24 & 41.62 & 50.71 & \raisebox{0.1ex}{\color{ForestGreen}{$\blacktriangle$}}~9.09 \\
        MBPP         & 47.35 & 53.83 & \raisebox{0.1ex}{\color{ForestGreen}{$\blacktriangle$}}~\phantom{a}6.48  & 44.76 & 53.04 & \raisebox{0.1ex}{\color{ForestGreen}{$\blacktriangle$}}~8.28 \\
        McEval Hard  & 12.47 & 26.08 & \raisebox{0.1ex}{\color{ForestGreen}{$\blacktriangle$}}~13.61 & 25.86 & 32.60 & \raisebox{0.1ex}{\color{ForestGreen}{$\blacktriangle$}}~6.74 \\
        \Xhline{2\arrayrulewidth}

        \rowcolor{lightGray}
        \multicolumn{7}{c}{\textbf{Qwen 3 8B}} \\
        \Xhline{1\arrayrulewidth}
        \rowcolor{veryLightGray}
        \textbf{} &
        \multicolumn{3}{c!{\vrule width 1pt}}{\textbf{Gleam}} &
        \multicolumn{3}{c}{\textbf{MoonBit}} \\
        \Xhline{1\arrayrulewidth}
        \textbf{Benchmark} & \makecell[c]{\textbf{PT}} & \makecell[c]{\textbf{Diff}} & \makecell[c]{$\Delta$} & \makecell[c]{\textbf{PT}} & \makecell[c]{\textbf{Diff}} & \makecell[c]{$\Delta$} \\
        \Xhline{2\arrayrulewidth}
        HumanEval    & 18.57 & 51.88 & \raisebox{0.1ex}{\color{ForestGreen}{$\blacktriangle$}}~33.31 & 36.82 & 44.42 & \raisebox{0.1ex}{\color{ForestGreen}{$\blacktriangle$}}~7.60 \\
        MBPP         & 23.63 & 50.48 & \raisebox{0.1ex}{\color{ForestGreen}{$\blacktriangle$}}~26.85 & 42.08 & 45.27 & \raisebox{0.1ex}{\color{ForestGreen}{$\blacktriangle$}}~3.19 \\
        McEval Hard  & 4.36  & 22.33 & \raisebox{0.1ex}{\color{ForestGreen}{$\blacktriangle$}}~17.97 & 19.87 & 19.82 & \raisebox{0.1ex}{\color{BrickRed}{$\blacktriangledown$}}~0.05 \\
    \end{tabularx}
    \end{tcolorbox}
\end{table}

\tabref{tab:diffResults} reports the results achieved on the no-resource languages via instruction transferring (see column ``Diff''), and their comparison against the best-performing approach highlighted in RQ$_2$, \ie the base model further pre-trained on the no-resource language (see column ``PT''). In what follows, we discuss the differences between the two approaches, while also highlighting the improvement achieved by instruction transferring with respect to other techniques experimented in RQ$_2$ (\tabref{tab:no_resource_approach}). Also in this case, we provide the full results of $passed_\%$ in the replication package \cite{replication}, highlighting relevant differences between the two metrics (\ie $pass@1$ and $passed_\%$) in the following.

Instruction transferring yields a significant improvement over the base model further pre-trained on the no-resource language, with $pass@1$ scores increasing by up to 33\% (Gleam, HumanEval, Qwen 3), and an average increase across benchmarks, models and languages of 12\%. All improvements are statistically significant (adjusted $p$-value $<$ 0.05, McNemar test), with an OR ranging between 1.21 and 10.95. There is only one case where the instruction transferring approach does not improve performance, namely $\langle$MoonBit, McEval-Hard, Qwen 3$\rangle$, although the difference is small (0.05\%) and not statistically significant (OR=1). Overall, we can safely state that instruction transferring significantly boosts the code generation capabilities of LLMs on no-resource languages. This boost is observed across different languages, benchmarks, and models. More importantly, it generalizes to models having different sizes (8B \emph{vs} 32B), being general-purpose (Qwen 3) or specialized on code (Qwen 2.5 Coder), and with or without reasoning capabilities  (Qwen 3 \emph{vs} Qwen 2.5 Coder).

It is worth noting that instruction transferring provides a more substantial improvement on Gleam than on MoonBit. Our hypothesis is that this is due to the fact that the improvement achieved by the further pre-training on MoonBit was already quite high, with an average increase in the $pass@1$ of 28\% with respect to the base model (see \tabref{tab:no_resource_approach}, Base, 0-shot). In contrast, the further pre-training on Gleam yielded a lower improvement (23\%), leaving more room for the instruction transferring to boost performance. 

When comparing the results against all LLMs and techniques evaluated in RQ$_2$ (\tabref{tab:no_resource_approach}), we observe that instruction transferring is the best approach in all cases. For instance, for the Gleam language, besides the ``further pre-training'' strategy shown in \tabref{tab:diffResults}, another high-performing approach is the fine-tuning of Qwen 3 32B Instruct, which achieved $pass@1$ scores of 24\% (HumanEval), 37\% (MBPP), and 4\% (McEval-Hard). Even so, the instruction transferring strategy applied on Qwen 2.5 Coder 32B Base achieves +33\%, +17\%, and +22\% improvements over these scores (measured in absolute terms), respectively. The same holds for MoonBit, where the second best-performing approach (after the further pre-training) is o3-mini complemented with a 5-shot in-context learning, which reached $pass@1$ scores of 39\% (HumanEval), 46\% (MBPP), and 12\% (McEval-Hard). Again, the instruction transferring approach applied on Qwen 2.5 Coder 32B Base achieves +11\%, +7\%, and +20\% improvements over these scores, respectively.

Lastly, we draw attention to an important finding from our experiments: instruction transferring applied to smaller models can result in outperforming larger and more expensive models. Indeed, Qwen 3 8B Base complemented with instruction transferring consistently outperforms Qwen 3 32B Instruct, whatever technique is applied on it. 

As compared to a fine-tuned Qwen 3 32B (best baseline from \tabref{tab:no_resource_approach}), Qwen 3 8B with instruction transferring obtained up to +28\% (measured in absolute terms) increments in $pass@1$ scores over the fine-tuned version of Qwen 3 32B Instruct (Gleam, HumanEval), with an average increase of 12\% across all benchmarks and languages. This finding highlights the exceptional efficiency of the instruction transferring technique, which allows smaller models to achieve superior performance compared to larger (4$\times$) alternatives. This is particularly relevant for practitioners working with computational constraints, as it suggests that strategic knowledge transfer can be more valuable than simply scaling model size.

While the results of the $passed_\%$ metric are consistent with what we discussed in terms of $pass@1$, we noticed one single difference worth being discussed.

Transitioning from pre-training to instruction transferring substantially reduces the total number of syntactic errors for the large model (32B): Qwen 2.5 exhibits a reduction of 24.1\% in syntactic errors on Gleam (from 752 to 571) and 73.5\% on MoonBit (from 645 to 171). Instead, on the small model (8B), besides an overall improvement in terms of performance, instruction transferring caused the failures to shift more towards syntactical errors as compared to the pre-trained model, both on Gleam (+13.3\%, from 460 to 521) and on MoonBit (+241.6\%, from 545 to 1,863). Clearly, such an increase of syntactical errors is accompanied by a stronger decrease of semantic errors, justifying the overall boost in performance. This divergent behavior between model sizes can be explained by differences in model capacity. For the larger model, instruction transferring effectively reinforces both syntactic competence and task-level understanding, leading to a substantial reduction of both syntactic and semantic errors. In contrast, the smaller model appears to reallocate its limited capacity toward improved instruction following and semantic reasoning. As a result, while overall performance improves, the remaining failures are more frequently attributable to syntactic issues. This shift also explains the only discrepancy we observed between the indications provided by $pass@1$ and those provided by $passed_\%$. Specifically, for Qwen 8B on the MoonBit translation of HumanEval, instruction transferring yields a higher $pass@1$ than pre-training (44.42 \emph{vs} 36.82), but a slightly lower $passed_\%$ (46.58 \emph{vs} 47.95). A plausible explanation is that syntactic errors, more frequent with instruction transferring, often prevent execution altogether, causing all tests for a task to fail rather than only a subset. Consequently, a model with more syntactic errors may still achieve a higher $pass@1$ by fully solving more tasks, while obtaining a lower $passed_\%$ because, on the tasks it fails, it more frequently fails all tests instead of only some of them.

\begin{tcolorbox}[title={\faLightbulbO \hspace{0.1cm} Answer to RQ$_3$}, colframe=black!90, colback=gray!10, boxrule=0.8pt, arc=4pt, left=1pt, right=1pt, top=1pt, bottom=1pt]
    Instruction transferring provides a strong boost in performance on no-resource languages as compared to the best-performing technique we experimented with (\ie further pre-training). Such a boost comes almost for free, and allows even smaller models to become competitive with larger ones.
\end{tcolorbox}

\section{Validity Discussion} \label{sec:threats}

\subsection{Experimental Procedure and Evaluated Techniques} 
We did not perform hyperparameter tuning of the experimented LLMs as this would have required a significant amount of computational resources. We used the default configurations suggested by the authors of the models. As for the number of training epochs (5), this was dictated by the monitoring of the loss function over training. As a double-check, we tested the models also after the first and third epochs, always getting worst results for the first epoch (indicating the need for more training) and quite similar results for the third (\ie at most a $\pm$3.4\% gap in $pass@1$ when considering all combinations of LLMs and languages).

For all experimented techniques we had to make choices. For in-context learning techniques, all prompts and scripts we used are publicly available \cite{replication}. Both in-context learning techniques we experiment with (\ie few-shot prompting and RAG)  dynamically select the most relevant code examples and documentation to be included in the prompt. However, this additional context may not always capture all language features required by the LLM to solve the task at hand. To assess the extent to which this could have impacted our findings, we implemented a third prompting strategy, consisting in a language-specific manual covering all information necessary to solve the tasks in our three benchmarks. To construct this manual, we first queried the documentation to identify the files most relevant to each benchmark task. This enabled us to reduce the full documentation to a representative subset of files relevant to at least one task. We then manually inspected the retrieved files to confirm their relevance to the corresponding benchmark tasks. Finally, we prompted GPT-4.1 to summarize these documents into a concise manual including short explanations and code examples for each required programming concept (the summarization prompt and the resulting manuals are available in our replication package \cite{replication}). The first two authors manually verified the accuracy of the generated Gleam and MoonBit manuals against the original documentation. The resulting manual can then be used as fixed prompt context for each code generation request submitted to an LLM.

We evaluated this strategy on Qwen 2.5 Coder 32B Instruct, as it is the only model for which we report results across all experimental settings (in-context learning, fine-tuning, continuous pre-training, and instruction transferring). As in the previous experiments, we performed 10 runs to account for the stochastic nature of LLM outputs. Before discussing the results, we note that this methodology, as applied here, is mainly useful for identifying an approximate upper bound on the performance achievable with in-context learning techniques. Indeed, the manual is effectively ``overfitted'' to a set of known code-generation tasks, namely those included in the benchmarks, which would not be known \emph{a priori} in a realistic usage scenario.

We obtained a $pass@1$ of 15.58\% on HumanEval, 25.07\% on MBPP, and 1.76\% of McEval-Hard for Gleam, and 32.47\% on HumanEval, 39.44\% on MBPP, and 6.61\% of McEval-Hard for MoonBit.
While results on the first two benchmarks are significantly better than any other in-context learning technique, and nearly on par with the fine-tuning baseline, they are still underperforming against our best technique (\ie instruction transferring). This is especially evident in the most challenging benchmark (\ie McEval-Hard), where it achieved a lower $pass@1$ (Gleam 1.76\%, MoonBit 6.61\%) compared to the model after instruction transferring (Gleam 26.08\%, MoonBit 32.60\%).

Finally, in RQ$_2$ we observed that LLMs mainly fail on no-resource languages due to syntactical errors. One possible approach to mitigate these errors is via constrained decoding techniques \cite{willard2023efficient, dong2025xgrammar}, which enforce language grammar rules at inference time. While we acknowledge that these techniques can boost open-weight models' performance without additional training, they do not take into account language APIs, semantic correctness, and  language-specific coding conventions. Therefore, we did not include these techniques in our study but we plan to experiment with them in future work.

\subsection{Experimental Assumptions}
A key assumption underlying our study is that the evaluated LLMs have not been exposed to Gleam or MoonBit code during training. However, the training corpora of the evaluated LLMs are not publicly characterized at a level that would allow us to determine whether Gleam or MoonBit code was present. For the open models considered in this work, we therefore cannot verify the absence of these languages from the training data. We found that neither Gleam nor MoonBit appeared among the 92 supported languages listed in the \texttt{README.md} file associated with the commit introducing Qwen 2.5 Coder \cite{qwen2-readme}. Similarly, the most recent version of \texttt{README.md} for Qwen 3 \cite{qwen3-readme} omits Gleam and MoonBit from the 358 supported programming languages. We interpret this only as evidence that these languages are not advertised as supported by the models, rather than as evidence about the contents of the training corpora. Nevertheless, the languages' timelines suggest that any such exposure was likely limited: Gleam reached version 1.0 only in March 2024, while MoonBit was still undergoing alpha testing in late 2023 and reached beta in June 2025.

\subsection{Data Quality}
One of the techniques we explore to improve LLM performance on no-resource languages is fine-tuning. This approach relies on the automated collection of $\langle$\emph{doc}, \emph{function}$\rangle$ pairs, where \emph{doc} denotes a natural-language description of the code to be implemented (provided as input to the LLM), and function represents the corresponding target code generation. In this process, the quality of the collected data is critical to the effectiveness of fine-tuning. To assess the quality of the fine-tuning datasets, we manually validated a random sample of 374 instances from the Gleam dataset and 333 instances from the MoonBit dataset. Both samples are statistically significant, ensuring a 95\% confidence level with a $\pm$5\% margin of error within each corresponding fine-tuning dataset. The first two authors manually validated the quality of the 707 summaries (\ie the \emph{doc} associated to each function) by following the same procedure used in works assessing the quality of manually \cite{Crupi:tse2025} or automatically-generated \cite{roy:fse2021} summaries. 
In particular, each summary has been independently assessed by the two evaluators across three dimensions: \emph{content adequacy}, \emph{conciseness}, and \emph{fluency \& understandability}. Each quality attribute has been assessed on a scale from 1 to 5 (the higher the better), using the guidelines defined by Crupi \etal \cite{Crupi:tse2025}. 

\begin{figure}[t]
    \centering
    \includegraphics[width=0.9\columnwidth]{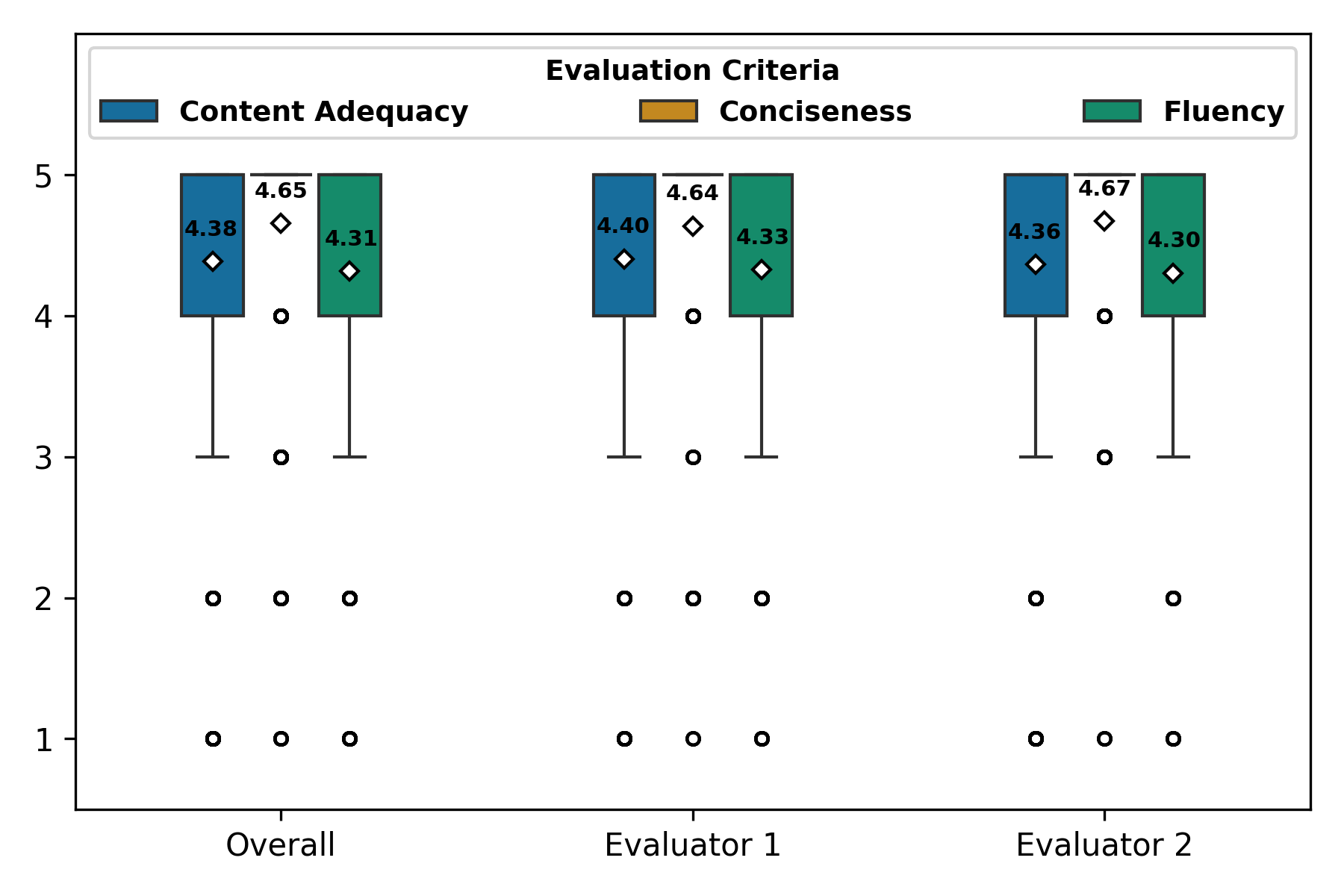}
    \caption{Distribution of the quality scores for the three evaluated criteria on the fine-tuning datasets.}
    \label{fig:boxplot}
\end{figure}

We computed the agreement among the human judges for the three quality aspects using the Krippendorff $\alpha$ \cite{krippendorff2011computing}, obtaining 0.69 for \emph{content adequacy} (substantial agreement), 0.48 for \emph{conciseness} (moderate), and 0.65 for \emph{fluency \& understandability} (substantial). \figref{fig:boxplot} shows the distribution of the three quality attributes when considering the evaluations provided by: (i)  both evaluators as a single distribution---overall; (ii) evaluator 1; and (iii) evaluator 2. As it can be seen, the median both overall as well as for each individual evaluator, is always 5, with average values never going below 4.3. These results give us confidence about the overall quality of the fine-tuning datasets we experimented with. Still, as for any dataset collected in the wild, we acknowledge the presence of low-quality instances which, however, represent the minority of the inspected instances (see \figref{fig:boxplot}). We also acknowledge a possible validity threat because the analysis was performed post-hoc by the first two authors, which may have introduced some bias in the evaluation. We tried to minimize such a bias by following the evaluation procedure described above.

Also, our findings could be influenced by errors made during the benchmarks' translations. To check the quality of the translated benchmarks, we randomly selected a set of instances for a further manual inspection. In particular, we created a statistically-significant sample (95\%$\pm$5\%) of translated coding tasks, stratified across the nine languages involved in our study. Our population is composed by a total of 3,061 translated coding tasks (\ie 227$\times$9 = 2,043 for the creation of McEval-Hard, 154$\times$2 = 308 for the translation of HumanEval in Gleam and MoonBit, and 355$\times$2 = 710 for the translation of MBPP in Gleam and MoonBit). Given such a population, the selected sample features 346 coding tasks, with 82 being Gleam, 82 being MoonBit, and the remaining 182 equally split across the remaining 7 languages (26 each). Basically, in such a sample we include more instances for more represented languages in the translated benchmarks (\ie Gleam and MoonBit). Then, the first two authors independently looked at each instance, comparing it with the original coding task from which it has been translated. Both authors had to answer two boolean questions: (i) is the prompt (\ie code description) of the translated coding task equivalent to that of the original task it stems from?; and (ii)~are the tests of the translated coding task equivalent to those of the original task it stems from? We also used this analysis to check for possible errors inherited by the source benchmarks (\eg misalignments between what required in the \texttt{docstring} and what tested in the test suite). There were 23 cases (6.6\%) in which the evaluators disagreed in the equivalence/correctness of the prompt and 14 (4\%) of the tests. We solved conflicts via open discussion. As a result, we found 14 prompts (4\%) and two tests (0.6\%) presenting issues. The two instances with incorrect test translations originate from task CPP/17 in McEval-Hard. Specifically, the original test suite validates the function robustness by passing binary and hexadecimal literals as arguments. In our R and Lua translations, however, we converted these numbers to decimals (\eg translating \textit{0b11011101111} to \textit{1775}). Although these translations are partially accurate (\eg Lua lacks native binary literals), they may not be considered as fully equivalent to the original tests. Regarding the 14 prompts we flagged as problematic: five errors were inherited issues from the McEval-Hard benchmarks that we did not spot while translating the original benchmark, such as doctests with incorrect return values; four were typos/textual deviations from the original prompt, such as missing a sentence; three related to the lack of type-related information, such as missing type hints in the function arguments; two were instead type-related errors (\eg in MoonBit one doctest reported ``\texttt{true}'' as the expected return value, while it should be ``\texttt{True}''). Overall, we detected a very low percentage of potential quality issues, which are inherent of any built benchmark, as also shown by recent work investigating the quality of code generation benchmarks \cite{Siddiq:scam2024,Wang:icse2026}. For example, Wang \etal \cite{Wang:icse2026} recently showed that test suites in SWE-bench have significant flaws, leading to inflated issue-resolution rates by 6.4\%, on average.

\subsection{Generalizability} 
We experimented with nine languages, six LLMs, and three benchmarks, our findings may not generalize to other settings. Also, the specific tasks we used to test the selected LLMs may not reflect the real usage scenarios in which programming languages are used. Indeed, some programming languages might be more frequently used in specific domains, such as high-performance computing, game development, or IoT. While this is not the case for MoonBit, which is described as a general-purpose language in their official website \cite{moonbitWebsite}, Gleam is mainly optimized for distributed systems and web development. To complement our analysis, future work should target domain-specific tasks for these programming languages.

Finally, our work does not aim at improving LLMs for Gleam and MoonBit specifically, but rather at studying the gap between no-resource languages and higher-resource ones, as well as proposing and evaluating techniques to improve LLMs in no-resource settings. In this sense, the implications of our study remain relevant even if future LLMs are trained on Gleam and MoonBit.
\section{Related Work} \label{sec:related}

In recent years, numerous code generation techniques \cite{starcoder,reflexion,wizardcoder,mapcoder} and benchmarks \cite{humaneval, mbpp, evalplus, bigcodebench} have been proposed, with most of them targeting popular programming languages, such as Python and Java. In this section, we focus on code generation benchmarks presented for low-resource programming languages and past attempts aiming at improving LLMs' performance in this scenario.

\textbf{Code Generation Benchmarks for Low-Resource Languages.} Several works proposed the translation of monolingual benchmarks to multiple programming languages, including low-resource ones~\cite{mbxp, babelcode, cassano:tse2023}. A notable example is \textsc{MultiPL-E} \cite{cassano:tse2023}, in which the authors translated \textsc{HumanEval} \cite{humaneval} and \textsc{MBPP} \cite{mbpp} to 24 languages. We exploit the \textsc{MultiPL-E} translated benchmarks to assess the performance of several LLMs on the low-resource programming languages subject of our study. Also, we translated those same benchmarks in Gleam and MoonBit (no-resource languages).  

Due to the raising concerns of data contamination between LLMs' training corpora and popular benchmarks, 
Chai \etal released \textsc{McEval} \cite{mceval}, a benchmark featuring 2k human-written code generation tasks. \textsc{McEval} supports 40 programming languages, with about 50 code instances each. We revised the original McEval benchmark to create a more challenging and robust version of it for low- and no-resource programming languages.

\textbf{Empirical Studies on Low-Resource Programming Languages.} There are several works tackling the code generation problem for low-resource languages, some of which have been explictly crafted for a given language, hardly being generalizable to other languages. For example, Kogler \etal \cite{kogler2025code} proposed the use of an intermediate DSL crafted as a JSON for a specific task and language of interest (\eg generating test specifications using the Balise Telegram Test Language). As also highlighted by the authors, their technique is not easily generalizable to other scenarios. In this section, we discuss techniques which, as the ones we experimented with, can be applied to any language. For a more complete overview of the area which includes the more specialized techniques, we invite the reader to check the recent survey by Joel \etal \cite{joel2024survey}.

Early attempts to improve performance on low-resource languages~\cite{ahmed:icse2022, chen:icpc2022, babelcode} deal with the shortage of training data by fine-tuning on multilingual datasets. For example, Chen \etal \cite{chen:icpc2022} found that training a DL model on programming languages similar to the niche ones of interest can boost performance on the latter. Our findings confirm, at least in part, what observed by Chen \etal Indeed, the LLMs used in our study have been pre-trained on a multitude of programming languages and worked well on some of the low-resource languages we experimented with. Instead, the multilingual pre-training did not help on the no-resource languages.

Cassano \etal \cite{cassano2023knowledge} proposed \textsc{MultiPL-T}, a framework to generate synthetic training data for low-resource languages. They used \textsc{MultiPL-T} to translate Python functions to multiple low-resource languages, using LLMs and language-specific compilers. They found that this technique can improve LLMs code generation performance on five low-resource languages. Paul \etal \cite{ircoder:acl2024} leveraged LLVM's intermediate representation \cite{llvm} to align code from popular and low-resource languages through a shared representation. They found that by continuously pre-training LLMs on pairs of source code and its intermediate representation, the code generation abilities of these models are greatly enhanced on niche languages. 

In our study, we did not include \textsc{MultiPL-T} and the approach by Paul \etal \cite{ircoder:acl2024} as baselines since the former cannot be applied in the context of no-resource languages, given the inability of LLMs to support the automated translation from high- to no-resource languages. For the latter, instead, there is no support of the LLVM Compiler Infrastructure for Gleam and MoonBit.

Other studies explored in-context learning techniques in the low-resource scenario.
Athiwaratkun \etal \cite{mbxp} investigated few-shot learning as an alternative method for teaching new languages to code models. Their experiments show that prepending some code examples on the original prompt can help a model to generate more accurate code. Dutta \etal \cite{rar:emnp2024} proposed \textsc{RAR}, a retrieval-augmented technique to guide model completion in low-resource languages. \textsc{RAR} uses a two-step retrieval process that relies on language documentation: it first retrieves relevant grammar definitions (\ie classes, methods, properties) given the code completion context and then extracts code examples from these. Their experiments show great improvements over existing baselines. Finally, Giagnorio \etal \cite{giagnorio:icpc2025} analyzed several in-context learning and fine-tuning strategies to improve models' code abilities in low-resource programming languages. Their findings reveal that providing more examples in the LLM context aids the model to generate better code, while fine-tuning techniques may actually reduce model performance. We used few-shot, RAG, and fine-tuning as baselines.

In a related line of research, Costa \etal recently proposed ModelMate \cite{modelmate}, an approach that fine-tunes pre-trained language models to provide editor assistance for textual domain-specific languages (DSLs) with little or no available training data. Their key idea is to increase the amount of training data by converting models written in a semantically compatible modeling language into the target DSL through a model-to-text transformation. This approach, however, relies on the assumption that such a transformation can be constructed for the target language. While this assumption is reasonable for DSLs with narrow, domain-constrained semantics, our work instead focuses on no-resource general-purpose programming languages, which exhibit broader syntax and semantics and are used for end-to-end program synthesis, making such assumptions less applicable.

\section{Conclusion and Future Work} \label{sec:conclusion}

We empirically evaluated the code generation capabilities of six state-of-the-art LLMs on high-, low-, and no-resource programming languages (\ie languages having abundant, little, and almost no training data, respectively). To run such an evaluation, we invested major effort in the creation of (i)~benchmarks for two no-resource languages (Gleam and MoonBit), and (ii) a more complex benchmark for the experimented high- and low-resource languages (\ie McEval-Hard). This required a total of $\sim$340 man hours. All our benchmarks are available to foster research in code generation \cite{replication}.

On McEval-Hard, LLMs achieved a $pass@1$ in the range of $\sim$59-89\% for high-resource languages, 27-84\% for low-resource, and  0-1\% for no-resource. We then investigated possible solutions to boost performance on no-resource languages, with the \textit{pre-training} of base models on the little data available resulting as the most effective technique, achieving a $pass@1$ up to 12\% on McEval-Hard for Gleam and 26\% for MoonBit. We also explored \textit{fine-tuning reuse} to transfer the instruction-tuned weights of an instruct model to a base model specialized (via pre-training) on the no-resource language. This approach considerably outperformed the other techniques, achieving a $pass@1$ of up to 26\% for Gleam and 33\% for MoonBit on McEval-Hard. If we compare these values to what originally observed in a zero-shot setting for the same languages (\ie $pass@1$ close to 0), our empirical investigation represents a starting point for companies interested in training and deploying their in-house AI-based coding assistant specialized on proprietary languages. 

Our future work will target the expansion of the current benchmarks to cover a more diverse set of real-world use cases in no-resource languages, like debugging, refactoring, or the generation of code aimed at addressing entire change requests. 

\section*{Acknowledgments}
Giagnorio and Bavota acknowledge the financial support of the Swiss National Science Foundation for the PARSED project (SNF Project No. 219294).
We also thank Louis Pilfold, creator of the Gleam language, for the help in verifying the correctness of Gleam benchmarks, and the entire Gleam and MoonBit communities for their support in the development of the benchmarks.

\bibliographystyle{IEEEtran}
\bibliography{main}

\end{document}